\documentclass[aps,preprint,onecolumn,12pt,nofootinbib]{revtex4}
\usepackage[utf8]{inputenc}
\usepackage[english]{babel}

\usepackage{graphicx}
\usepackage{subcaption}
\usepackage{caption}
\usepackage{tikz}
\usetikzlibrary{quotes,angles,arrows,decorations.markings}

\usepackage{amsmath,amssymb,amsfonts,mathrsfs}
\usepackage{bm,bbm,dsfont,braket,tensor,esint}
\usepackage{cancel,booktabs,multirow,array,makecell}
\newcolumntype{C}{>{\centering\arraybackslash}p{1.8cm}}

\usepackage{color,xcolor}
\usepackage{float}
\usepackage{lipsum}
\usepackage{enumerate}
\usepackage{changes}
\usepackage{ulem}
\setaddedmarkup{\textcolor{red}{#1}}
\setdeletedmarkup{\textcolor{gray}{\sout{#1}}}

\usepackage{hyperref}
\hypersetup{
    colorlinks,
    breaklinks,
    citecolor=blue,
    filecolor=green,
    linkcolor=purple,
    urlcolor=red,
    citecolor=[rgb]{0,0.0,1.0},
    urlcolor=[rgb]{0.0,0.0,1.0},
    linkcolor=[rgb]{0,0.5,0.9}
}

\usepackage{slashed}

\definecolor{lime}{HTML}{A6CE39}
\DeclareRobustCommand{\orcidicon}{%
    \begin{tikzpicture}
    \draw[lime, fill=lime] (0,0) 
    circle [radius=0.16] 
    node[white] {{\fontfamily{qag}\selectfont \tiny ID}};
    \draw[white, fill=white] (-0.0625,0.095) 
    circle [radius=0.007];
    \end{tikzpicture}
    \hspace{-2mm}
}

\foreach \x in {A, ..., Z}{%
    \expandafter\xdef\csname orcid\x\endcsname{\noexpand\href{https://orcid.org/\csname orcidauthor\x\endcsname}{\noexpand\orcidicon}}
}

\begin{document}

\title{The Barrow entropies in the thermodynamics of high-dimensional Gauss-Bonnet black holes}

\author{Yuxuan Shi\orcidA{}}
\email{shiyx2280771974@gmail.com}
\affiliation{School of Physics, East China University of Science and Technology, Shanghai 200237, China}

\author{Hongbo Cheng}
\email{hbcheng@ecust.edu.cn}
\affiliation{School of Physics, East China University of Science and Technology, Shanghai 200237, China}
\affiliation{The Shanghai Key Laboratory of Astrophysics, Shanghai 200234, China}

\begin{abstract}
We study the thermodynamics of $D$-dimensional Gauss-Bonnet black holes with Barrow entropy. It is found that the Gauss-Bonnet coupling and the Barrow factor revise the thermodynamic variables such as event horizon, Hawking temperature, entropy and heat capacity. It is interesting that the larger five-dimensional black holes exist stably and the smaller ones evaporate to disappear owing to the nature of heat capacities amended by the coupling and factor. The discussions exhibit that the $D$-dimensional black holes with $D=6, 7$ set free all of their energy to vanish because of the minus heat capacities as functions of Hawking temperature although the extra term and the fractal power bring about their revisions on the functions, but they cannot change the heat capacity signs, so they also cannot change the fate of six- or seven-dimensional black holes. 
\end{abstract}

\maketitle

%\tableofcontents

\section{Introduction}

Investigating the physical world with more than four dimensions is well-motivated, largely because high-dimensional spacetime provides the necessary ground for establishing more comprehensive models. In such higher-dimensional scenarios, the Gauss-Bonnet term naturally appears as the leading order correction in Lanczos-Lovelock gravity. It is worth noting that this type of gravity becomes topologically trivial (or non-dynamical) if the dimensionality is restricted to four, which necessitates considering $D>4$. The Gauss-Bonnet term is not just the lowest order correction to Lanczos’ theory but also constitutes a special sector of Lovelock’s theory of gravitation \cite{lanczos1932elektromagnetismus,lanczos1938remarkable,lovelock1971einstein,lovelock1972four}. Furthermore, Einstein-Gauss-Bonnet gravity attracts considerable interest due to its roots in string theory. In fact, this generalized gravity involves a dominant quantum correction to classical general relativity, and the so-called Gauss-Bonnet coupling arises naturally in the low-energy limit of heterotic superstring theory \cite{abdalla2005scalar,gleiser2005linear,sahabandu2006thermodynamics,dominguez2006radiating,maeda2006kaluza,sadeghi2014strong}. Being quadratic in curvature, the Gauss-Bonnet term in the Lagrangian certainly helps in regularizing the spacetime structure \cite{myers1988black,jacobson1993black,kothawala2008gravitational}. Over the years, quite a bit of effort has been devoted to the Lanczos-Lovelock model. For instance, gravitational lensing governed by Gauss-Bonnet gravity in the strong field limit has been analyzed \cite{sadeghi2014strong,man2014time}, and extensive research has been performed on the thermodynamic properties of black holes within this framework \cite{mukhopadhyay2006holography,paranjape2006thermodynamic}. More recently, black holes with fractal structures—viewed as manifestations of quantum gravity spacetime foam—have been discussed via surface gravity analysis \cite{abreu2025surface}.

Black hole thermodynamics has been a central pillar of theoretical physics for over half a century. In this context, black holes are treated as genuine thermodynamic systems characterized by temperature, entropy, and other standard potentials \cite{chandrasekhar1998mathematical,wald2001thermodynamics,johnson2020microscopic,chandran2020one,hamil2021effect,hendi2021physical,shalaby2021non}. The macroscopic behavior of these compact objects is governed by the four laws of black hole mechanics \cite{chandrasekhar1998mathematical,wald2001thermodynamics,johnson2020microscopic,chandran2020one,hamil2021effect,hendi2021physical,shalaby2021non}. Among these properties, entropy plays a pivotal role. It is well established that black holes emit thermal radiation due to quantum mechanical effects \cite{hawking1974black,hawking1975particle}, a phenomenon closely tied to Bekenstein's famous conjecture connecting entropy to the horizon area \cite{bekenstein1973black,bekenstein1973nuovo,bardeen1973four}. Since entropy is intrinsically linked to the horizon geometry, any modification to the entropy formula inevitably alters other thermodynamic quantities and, consequently, the evolution of the black hole \cite{chandrasekhar1998mathematical,wald2001thermodynamics,hawking1974black,hawking1975particle,bekenstein1973black,bekenstein1973nuovo,bardeen1973four}.

Crucially, it has been suggested that quantum fluctuations might induce a fractal structure in spacetime \cite{abreu2020barrow1,abreu2020barrow2,abreu2020barrow3,abreu2020thermal,saridakis2020modified,anagnostopoulos2020observational,di2022barrow,di2022sign}. To capture this feature, Barrow proposed a model where quantum gravitational effects roughen the static, spherically symmetric horizon into a fractal structure \cite{barrow2020area}. This deformation requires a reformulation of the standard area law. Specifically, the black hole entropy is redefined using an exponent $\Delta$, which quantifies the degree of fractality induced by quantum gravity \cite{barrow2020area}. This Barrow parameter is typically constrained to the range $0\le\Delta\le1$, and it is important to note that the standard Bekenstein-Hawking entropy is seamlessly recovered in the limit $\Delta=0$ \cite{hawking1974black,hawking1975particle,bekenstein1973black,bekenstein1973nuovo,bardeen1973four,barrow2020area}. Recent studies on Barrow-corrected Schwarzschild black holes—examining quantities like Helmholtz free energy and heat capacity—have shown that these corrections tend to suppress Hawking radiation \cite{petridis2023barrow}. Furthermore, other variations, such as logarithmically corrected Barrow entropy motivated by loop quantum gravity, have been found to significantly modify thermodynamic stability and black hole lifetimes \cite{capozziello2025barrow}.

It is therefore crucial to investigate the thermodynamics of higher-dimensional black holes that incorporate both Gauss-Bonnet corrections and Barrow-type fractal structures. While black holes within the framework of higher-dimensional Gauss-Bonnet gravity have been explored from various angles \cite{abdalla2005scalar,gleiser2005linear,sahabandu2006thermodynamics,dominguez2006radiating,maeda2006kaluza,sadeghi2014strong,myers1988black,jacobson1993black,kothawala2008gravitational,man2014time,mukhopadhyay2006holography,paranjape2006thermodynamic,abreu2025surface}, the influence of quantum gravity on their thermodynamic nature cannot be overlooked. Standard quantities such as the event horizon, Hawking temperature, and entropy describe the evolution and stability of these objects \cite{chandrasekhar1998mathematical,wald2001thermodynamics,johnson2020microscopic,chandran2020one,hamil2021effect,hendi2021physical,shalaby2021non}. However, Barrow's introduction of a fractal-like event horizon leads to generalized entropies that significantly alter these thermodynamic characteristics \cite{abreu2020barrow1,abreu2020barrow2,abreu2020barrow3,abreu2020thermal,saridakis2020modified,anagnostopoulos2020observational,di2022barrow,di2022sign,barrow2020area,petridis2023barrow,capozziello2025barrow}. Although recent surface gravity analysis has touched upon the five-dimensional case \cite{abreu2025surface}, a comprehensive scrutiny of the relationship between the evolution of generic $D$-dimensional Gauss-Bonnet black holes and these quantum gravitational effects is still necessary. We are particularly interested in how the interplay between the Barrow parameter and the Gauss-Bonnet coupling dictates the stability of black holes across different dimensions. In this work, we consider $D$-dimensional Gauss-Bonnet black holes, derive their corresponding Barrow entropies and heat capacities, and finally, provide a detailed discussion on their thermodynamic properties.

%%%%%%%%%%%%%%%%%%%%%%%%%%%%%%%%%%%%%%%%%%%%%%%%%%%%%%%%%%%%%%%%
%%%%%%%%%%%%%%%%%%%%%%%%%%%%%%%%%%%%%%%%%%%%%%%%%%%%%%%%%%%%%%%%
%%%%%%%%%%%%%%%%%%%%%%%%%%%%%%%%%%%%%%%%%%%%%%%%%%%%%%%%%%%%%%%%

\section{Theoretical Framework and Thermodynamic Quantities}

We start by examining the metric of a higher-dimensional Gauss-Bonnet black hole to gain insight into the connection between the gravitational horizon and thermodynamics. The Lagrangian for gravity in $D$-dimensional spacetime, composed of the first two terms of Lovelock theory (which includes the quadratic curvature parts), is written as follows \cite{lanczos1932elektromagnetismus,lanczos1938remarkable,lovelock1971einstein,lovelock1972four,myers1988black,jacobson1993black}:
\begin{align}
\label{lagrangian}
\mathcal{L}_{G}&=\dfrac{1}{16\pi}\Big[R+\alpha_{GB}\Big(R^{2}-4R_{MN}R^{MN}+R_{MNPQ}R^{MNPQ}\Big)\Big],
\end{align}
where $R$, $R_{MN}$, and $R_{MNPQ}$ are the $D$-dimensional Ricci scalar, Ricci tensor, and Riemann curvature tensor, respectively. The parameter $\alpha_{GB}$ represents the Gauss-Bonnet coupling, which relates to the inverse string tension \cite{zwiebach1985curvature,boulware1985string}. From the Lagrangian \eqref{lagrangian}, the field equations yield static and spherically symmetric black hole solutions, which can be presented as \cite{myers1988black,jacobson1993black,abreu2025surface}
\begin{align}
\label{metric1}
\mathrm{d}s^{2}=f(r)\mathrm{d}t^{2}-\dfrac{\mathrm{d}r^{2}}{f(r)}-r^{2}\mathrm{d}\Omega_{D-2}^{2},
\end{align}
where the metric function takes the form
\begin{align}
\label{fr1}
f(r)=1+\dfrac{r^{2}}{2\alpha}\left(1-\sqrt{1+\dfrac{4\alpha\omega}{r^{D-1}}}\right),
\end{align}
with $\alpha=(D-3)(D-4)\alpha_{GB}$. The ADM mass $M$ is related to the parameter $\omega$ via $\omega=(16\pi M)/[(D-2)V_{D-2}]$ \cite{abreu2025surface}, where $V_{D-2}$ represents the volume of a unit $(D-2)$-sphere. The last term in metric \eqref{metric1} is simply the angular part. Defining the horizon radius $r_+$ as the root of $f(r_{+})=0$, we can express the black hole's mass as \cite{abreu2025surface}
\begin{align}
\label{Mass_GB}
M_{GB}=\dfrac{V_{D-2}}{16\pi}(D-2)r_{+}^{D-5}\left(r_+^2+\alpha\right),
\end{align}
which satisfies the condition:
\begin{align}
\label{fr=0}
r_{+}^{D-3}+\alpha r_{+}^{D-5}-\omega=0.
\end{align}

Having established the background geometry, we briefly review the standard thermodynamics. The Hawking temperature of the higher-dimensional Gauss-Bonnet black hole is given by \cite{abreu2025surface,hawking1974black,hawking1975particle}
\begin{align}
\label{temperature_H}
T_{GB}=\left.\dfrac{1}{4\pi}\dfrac{\mathrm{d}f(r)}{\mathrm{d}r}\right|_{r=r_+}=\dfrac{(D-3)r_+^2+(D-5)\alpha}{4\pi r_+\left(r_+^2+2\alpha\right)},
\end{align}
which depends on the horizon radius $r_{+}$ determined by Eq.\eqref{fr=0}. The standard entropy for high-dimensional black holes in Gauss-Bonnet gravity obeys the area law with a curvature correction \cite{abreu2025surface,bekenstein1973black,bekenstein1973nuovo,bardeen1973four}:
\begin{align}
\label{entropy_GB}
S_{GB}=\int\dfrac{1}{T}\dfrac{\partial M}{\partial r}\mathrm{d}r=\dfrac{A_{D+}}{4}\left(1+\dfrac{2\alpha}{r_+^2}\dfrac{D-2}{D-4}\right),
\end{align}
where $A_{D+}=V_{D-2}r_{+}^{D-2}$ is the horizon surface area.

Now, we turn to the quantum gravitational corrections. It is widely believed that quantum gravity effects may alter the structure of spacetime at the Planck scale. One compelling approach suggests that the black hole horizon might exhibit a fractal structure, similar to a "spacetime foam," rather than being perfectly smooth \cite{barrow2020area}. Conceptually, this scheme involves adding smaller spheres to the surface in an iterative process, potentially leading to a finite volume but an intricate, infinite area \cite{barrow2020area}. Following the framework suggested in Refs.\cite{barrow2020area,petridis2023barrow,falconer2013fractal}, if miniature spheres with a hierarchical radius $r_{n+1}=\lambda r_{n}$ are stacked over $N$ steps, the resulting area and volume scale as:
\begin{align}
\label{A_inf}
A_{\infty}=\sum_{n=0}^{\infty}N^nV_{D-2}\left(\lambda^{n}r_{+}\right)^{D-2}=\dfrac{A_{D+}}{1-N\lambda^{D-2}},
\end{align}
and
\begin{align}
\label{V_inf}
V_{\infty}=\sum_{n=0}^{\infty}N^n\dfrac{V_{D-2}}{D-1}\left(\lambda^{n}r_{+}\right)^{D-1}=\dfrac{V_{D+}}{1-N\lambda^{D-1}},
\end{align}
respectively. Here, $V_{D+}=V_{D-2}r_{+}^{D-1}/(D-1)$. From Eq.\eqref{A_inf}, we see that if $N\lambda^{D-2}\to 1$, the area $A_{\infty}$ diverges. In practice, there exists a packing bound $N\leq\pi(1+1/\lambda)$ \cite{petridis2023barrow}. Based on Barrow's fractal geometry argument, the effective surface area scales as \cite{barrow2020area}:
\begin{align}
\label{V_D+}
\tilde{A}_{D}\varpropto r^{D-2+\Delta},
\end{align}
where the Barrow parameter $\Delta$ lies in the range $0\leq\Delta\leq1$. This formulation provides a modified ratio of the black hole's surface area to the Planck area \cite{barrow2020area}:
\begin{align}
\label{ratio}
\dfrac{\tilde{A}_{\infty}}{\tilde{A}_{p\infty}}=\left(\dfrac{A_{D+}}{A_p}\right)^{1+\frac{\Delta}{D-2}},
\end{align}
where $A_{p}=V_{D-2}\ell_{p}^{D-2}$ is the standard Planck area (with $\ell_{p}=\sqrt{G}\sim1$), and $A_{p\infty}$ represents the Planck area within the fractal geometry frame. Comparing Eq.\eqref{A_inf} with the ratio in Eq.\eqref{ratio}, we obtain the relation \cite{barrow2020area}:
\begin{align}
\label{ratio_new}
\dfrac{A_{D+}/A_p}{1-N\lambda^{D-2}}=\left(\dfrac{A_{D+}}{A_p}\right)^{1+\frac{\Delta}{D-2}}.
\end{align}
This expression can be rearranged to define the Barrow parameter explicitly \cite{barrow2020area}:
\begin{align}
\label{barrow_para}
\Delta=-\dfrac{(D-2)\ln\left(1-N\lambda^{D-2}\right)}{\ln\left(\frac{A_{D+}}{A_p}\right)}.
\end{align}
Clearly, $\Delta=0$ corresponds to the smooth limit where $N=0$.

Using the relation $A_{D+}/A_{p}=r_{+}^{D-2}$, the standard Gauss-Bonnet entropy (Eq.\eqref{entropy_GB}) can be rewritten in terms of dimensionless area ratios as \cite{abreu2025surface,bekenstein1973black,bekenstein1973nuovo,bardeen1973four}:
\begin{align}
\label{entropy_2}
S_{GB}=\dfrac{V_{D-2}}{4}\dfrac{A_{D+}}{A_p}+\dfrac{V_{D-2}\alpha}{2}\dfrac{D-2}{D-4}\left(\dfrac{A_{D+}}{A_p}\right)^{\frac{D-4}{D-2}}.
\end{align}
Following the logic in Refs.\cite{barrow2020area,petridis2023barrow,falconer2013fractal}, we generalize this entropy by replacing the standard area ratio with the fractal one defined in Eq.\eqref{ratio}. The Barrow-corrected Gauss-Bonnet entropy thus becomes:
\begin{align}
\label{entropy_GBB_1}
S_{GBB}=\dfrac{V_{D-2}}{4}\left(\dfrac{\tilde{A}_{\infty}}{\tilde{A}_{p\infty}}\right)+\dfrac{V_{D-2}\alpha}{2}\dfrac{D-2}{D-4}\left(\dfrac{\tilde{A}_{\infty}}{\tilde{A}_{p\infty}}\right)^{\frac{D-4}{D-2}}.
\end{align}
Substituting Eq.\eqref{ratio} into Eq.\eqref{entropy_GBB_1}, we arrive at the explicit form of the Gauss-Bonnet-Barrow entropy:
\begin{align}
\label{entropy_GBB_2}
S_{GBB}=\dfrac{V_{D-2}}{4}r_{+}^{D-2+\Delta}+\dfrac{V_{D-2}\alpha}{2}\dfrac{D-2}{D-4}r_{+}^{\frac{(D-4)(D-2+\Delta)}{D-2}}.
\end{align}
It is obvious that the geometric structure of the fractal horizon is quantified by the exponent containing the factor $\Delta\in[0, 1]$.

It is worth noting that these Barrow-corrected black holes still satisfy the first law of thermodynamics \cite{barrow2020area,jardim2012thermodynamics}:
\begin{align}
\label{1st_law}
\dfrac{1}{T_{GBB}}=\dfrac{\partial S_{GBB}}{\partial M_{GB}}=\frac{V_{D-2}}{4}(D+\Delta-2)r_+^{D-5}\left(2\alpha r_+^{\frac{D-4}{D-2}\Delta}+r_+^{\Delta +2}\right)\dfrac{\mathrm{d}r_+}{\mathrm{d}M_{GB}}.
\end{align}
Using the black hole mass from Eq.\eqref{Mass_GB} and the Barrow entropy from Eq.\eqref{entropy_GBB_2}, the first law yields the modified Hawking temperature \cite{barrow2020area,jardim2012thermodynamics}:
\begin{align}
\label{temperature_GBB}
T_{GBB}=\frac{(D-2)\left[\alpha(D-5)+(D-3)r_+^2\right]}{4\pi r_+(D+\Delta-2)\left(2\alpha r_+^{\frac{D-4}{D-2}\Delta}+r_+^{\Delta+2}\right)}.
\end{align}
This result generalizes the findings of Ref.\cite{abreu2025surface} to include Barrow corrections. It is manifest that in the limit $\Delta \to 0$, we recover the standard Gauss-Bonnet temperature $T_{GB}$ as given in Eq.\eqref{temperature_H}.

Finally, to address the local thermodynamic stability and phase structure of these black holes, we define the heat capacity as \cite{chandrasekhar1998mathematical,hawking1974black,hawking1975particle,bekenstein1973nuovo,bekenstein1973black}:
\begin{align}
\label{capacity}
C_{V,GBB}=T_{GBB}\dfrac{\partial S_{GBB}}{\partial T_{GBB}}.
\end{align}
Given the algebraic complexity of the explicit expression for $C_{V,GBB}$, we relegate the full formula to Appendix \ref{App.A}. Detailed numerical analysis regarding its physical behavior and the associated stability criteria will be presented in the next section.

%%%%%%%%%%%%%%%%%%%%%%%%%%%%%%%%%%%%%%%%%%%%%%%%%%%%%%%%%%%%%%%%
%%%%%%%%%%%%%%%%%%%%%%%%%%%%%%%%%%%%%%%%%%%%%%%%%%%%%%%%%%%%%%%%
%%%%%%%%%%%%%%%%%%%%%%%%%%%%%%%%%%%%%%%%%%%%%%%%%%%%%%%%%%%%%%%%

\section{Thermodynamic Stability and Phase Transitions}

In this section, we perform a detailed analysis of the thermodynamic stability and phase structure of $D$-dimensional Gauss-Bonnet black holes modified by Barrow entropy. Based on the analytical expressions derived in the previous section, we numerically investigate the behaviors of key thermodynamic quantities, including the modified Hawking temperature, the entropy-temperature relation, and the heat capacity. Our primary goal is to elucidate the effects between the Barrow fractal parameter $\Delta$ and the Gauss-Bonnet coupling $\alpha$, and to distinguish the stability characteristics across different spacetime dimensions.

\subsection{Behavior of Hawking Temperature}

\begin{figure}[tp]
\centering
\includegraphics[width=0.85\textwidth]{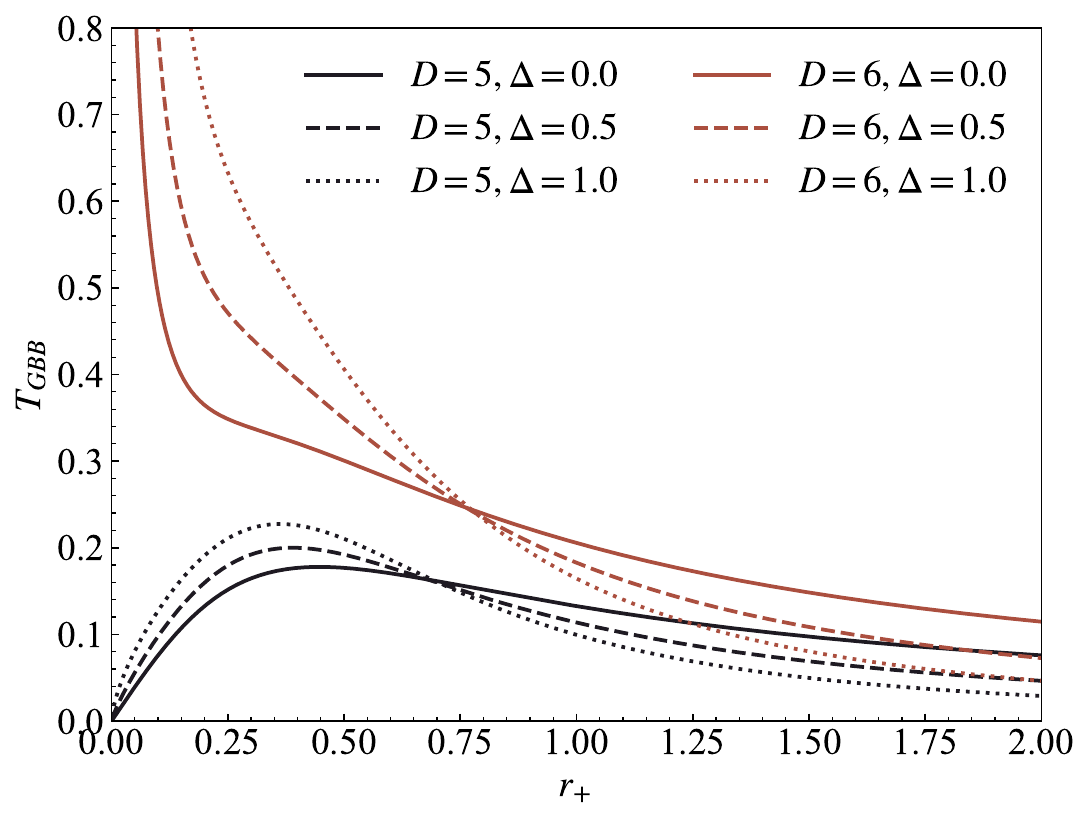}
\caption{The Hawking temperature $T_{GBB}$ as a function of the horizon radius $r_+$ for a fixed Gauss-Bonnet coupling $\alpha=0.1$. The black curves represent $D=5$ cases, while the red curves represent $D=6$. Different line styles correspond to different Barrow parameters: $\Delta=0.0$ (solid), $\Delta=0.5$ (dashed), and $\Delta=1.0$ (dotted).}
\label{fig:T_fix_alpha}
\end{figure}

\begin{figure}[tp]
\centering
\includegraphics[width=0.85\textwidth]{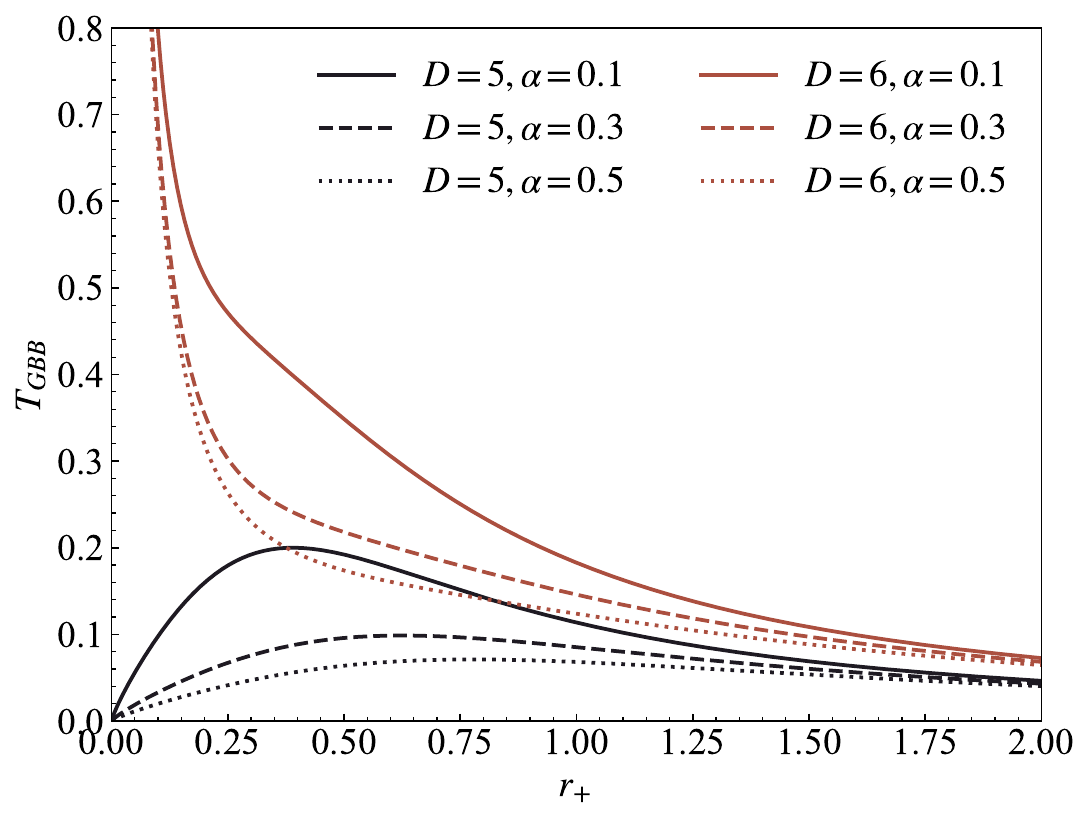}
\caption{The Hawking temperature $T_{GBB}$ as a function of $r_+$ for a fixed Barrow parameter $\Delta=0.5$. The black curves ($D=5$) and red curves ($D=6$) show the effect of varying the Gauss-Bonnet coupling: $\alpha=0.1$ (solid), $\alpha=0.3$ (dashed), and $\alpha=0.5$ (dotted).}
\label{fig:T_fix_delta}
\end{figure}

The behavior of the Hawking temperature as a function of the event horizon is illustrated graphically in Fig.\ref{fig:T_fix_alpha} and Fig.\ref{fig:T_fix_delta}. By observing the five-dimensional cases (the black curves), we find that the shapes of the temperature curves are qualitatively similar, featuring a characteristic peak. However, the quantitative behaviors are quite sensitive to the parameters. Specifically, in Fig.\ref{fig:T_fix_alpha}, we can see that a larger fractal factor $\Delta$ tends to push the temperature peak to a higher value and a smaller horizon radius. In contrast, as shown in Fig.\ref{fig:T_fix_delta}, the Gauss-Bonnet influence clearly lowers the temperature profile, acting as a cooling factor.

Things are different when we move to the six-dimensional spacetime (the red curves). Here, the temperature is strictly a decreasing function of the horizon $r_{+}$. The Barrow correction $\Delta$ lifts the entire curve, increasing the temperature, whereas the Gauss-Bonnet term suppresses it. We also checked the seven-dimensional case, and the results are essentially the same as those in six dimensions. For completeness, the corresponding temperature profiles for $D=7$ are presented in Appendix \ref{App.B} (see Fig.\ref{fig:Tr_D7_combined}).

It is interesting to note that while the Barrow parameter $\Delta$ and the Gauss-Bonnet coupling $\alpha$ compete with each other—one raising the temperature and the other lowering it—they essentially just adjust the magnitude. They do not drastically change the fundamental profile of the Hawking temperature curves, which is primarily determined by the spacetime dimension $D$.

\subsection{Entropy-Temperature Relation and Stability}

\begin{figure}[tp]
\centering
\includegraphics[width=0.85\textwidth]{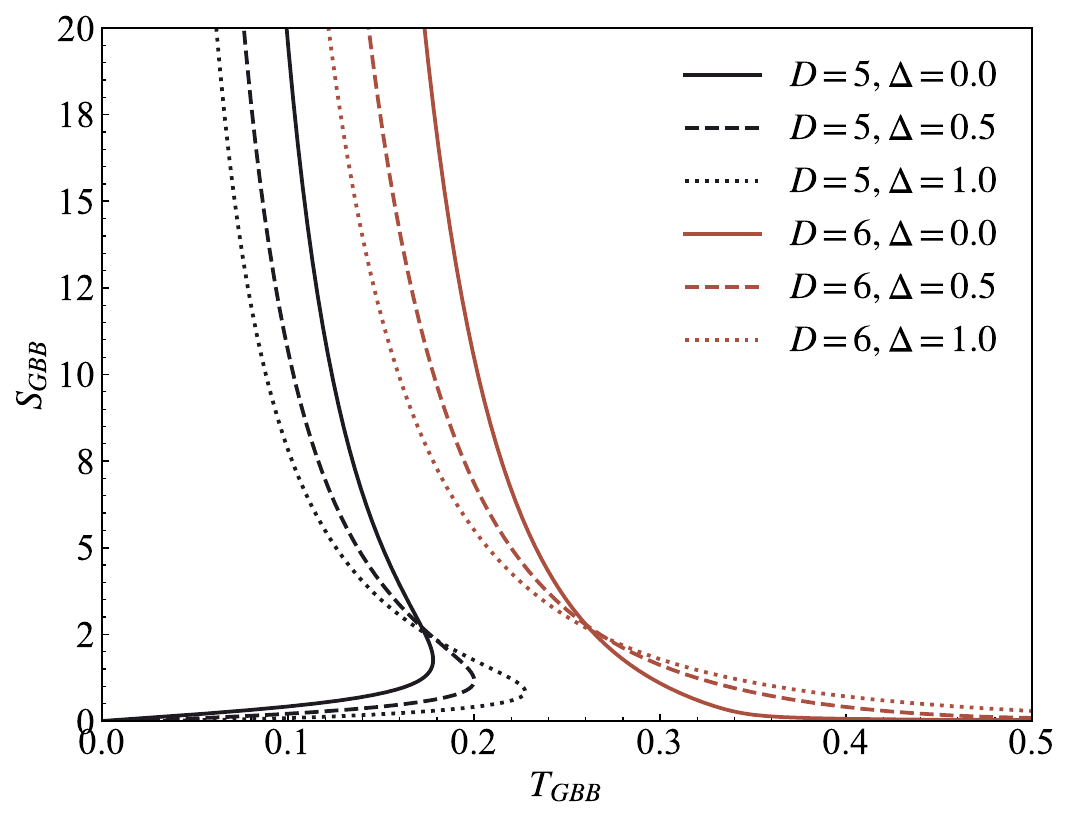}
\caption{The Entropy-Temperature $S_{GBB}-T_{GBB}$ diagrams for fixed $\alpha=0.1$. The presence of a cusp in the $D=5$ curves (black) indicates a phase transition, while the $D=6$ curves (red) show monotonic instability.}
\label{fig:ST_fix_alpha}
\end{figure}

\begin{figure}[tp]
\centering
\includegraphics[width=0.85\textwidth]{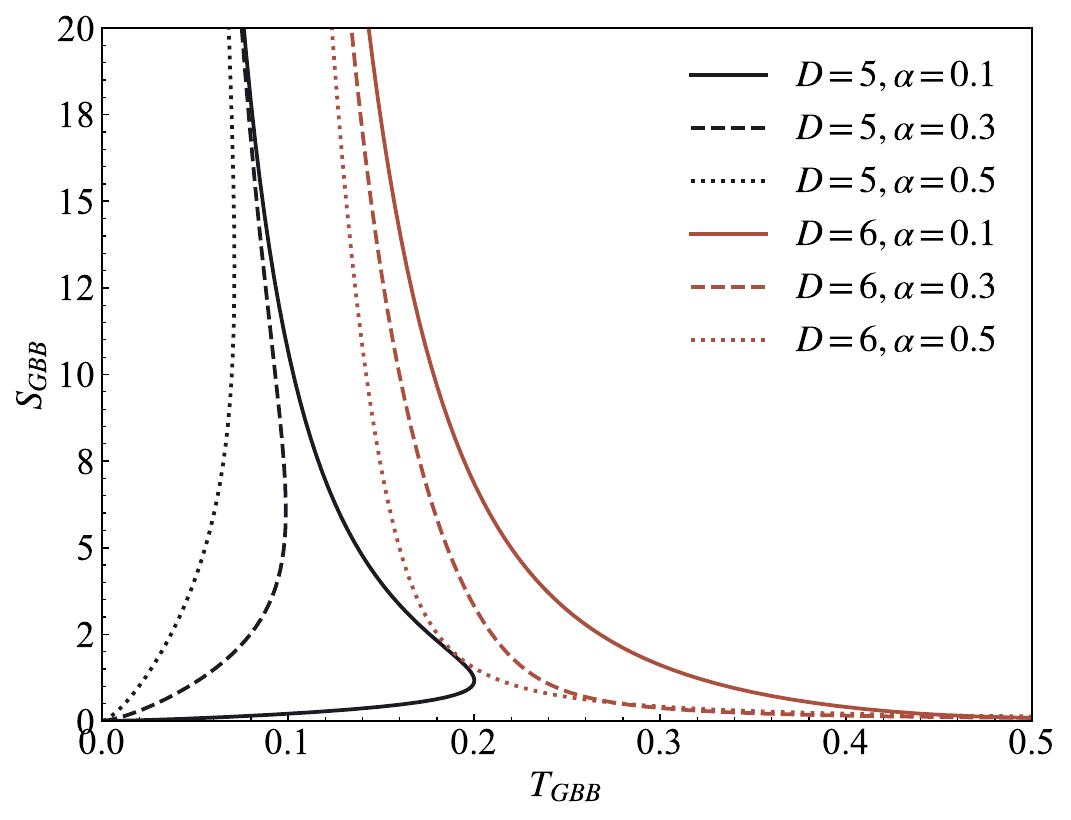}
\caption{The $S_{GBB}-T_{GBB}$ diagrams for fixed $\Delta=0.5$ with varying Gauss-Bonnet coupling $\alpha$. Larger $\alpha$ shifts the critical point to lower temperatures.}
\label{fig:ST_fix_delta}
\end{figure}

It is straightforward to verify that Eq.\eqref{entropy_GBB_2} reduces to the standard Gauss-Bonnet entropy $S_{GB}$ in Eq.\eqref{entropy_GB} when we take the limit $\Delta \to 0$. Since the exponents of $r_{+}$ in Eq.\eqref{entropy_GBB_2} are positive, the Barrow entropy naturally increases with the black hole size. To better understand the thermodynamic behavior, we plot the Barrow entropy $S_{GBB}$ against the Hawking temperature $T_{GBB}$ using the combined results of Eq.\eqref{entropy_GBB_2} and Eq.\eqref{temperature_GBB}. The resulting diagrams are displayed in Fig.\ref{fig:ST_fix_alpha} and Fig.\ref{fig:ST_fix_delta}.

Let's first look at the 5D case (the black curves). The general dependence of entropy on temperature is similar across these figures, featuring a distinct "cusp" structure. This cusp represents a critical point where the slope of the entropy curve, $\partial S_{GBB}/\partial T_{GBB}$, diverges. The lower branch of the curve (corresponding to smaller entropies) has a positive slope, indicating thermodynamic stability. Conversely, the upper branch (larger entropies) exhibits a negative slope, signaling instability.

Regarding the phase transition in $D=5$, our plots reveal an interesting competition between the parameters. As seen in Fig.\ref{fig:ST_fix_alpha}, increasing the fractal parameter $\Delta$ actually shifts the critical point to a higher temperature. On the other hand, Fig.\ref{fig:ST_fix_delta} shows that the Gauss-Bonnet coupling $\alpha$ acts to lower the critical temperature.

The situation is much simpler for higher dimensions ($D=6$, red curves). As shown in the right panels of the figures, the entropy profile resembles that of the seven-dimensional case, which refer to Fig.\ref{fig:ST_D7_combined} in Appendix \ref{App.B} for details. In these dimensions, the entropy is strictly a decreasing function of the corrected temperature. Neither the fractal structure nor the Gauss-Bonnet term changes the fundamental shape of the curves; they merely adjust the magnitude. Consequently, the slope of the Barrow-type entropy curves remains negative throughout, implying that these black holes are intrinsically unstable.

\subsection{Local Thermodynamic Stability Analysis}

\begin{figure}[tp]
\centering
\includegraphics[width=0.8\textwidth]{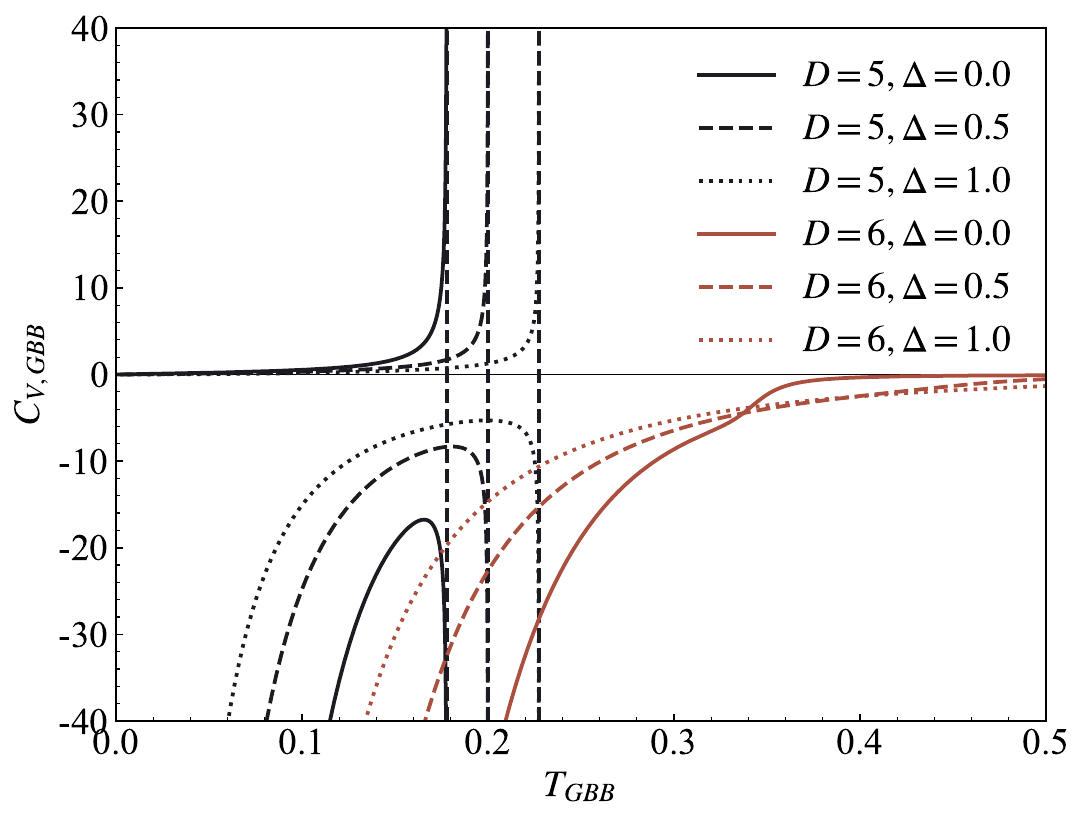}
\caption{The heat capacity $C_{V,GBB}$ versus Hawking temperature for fixed $\alpha=0.1$. The vertical dashed lines indicate the divergence of heat capacity for $D=5$ cases, marking the phase transition point. Note that for $D=6$ (red curves), the heat capacity is always negative.}
\label{fig:CV_fix_alpha}
\end{figure}

\begin{figure}[tp]
\centering
\includegraphics[width=0.8\textwidth]{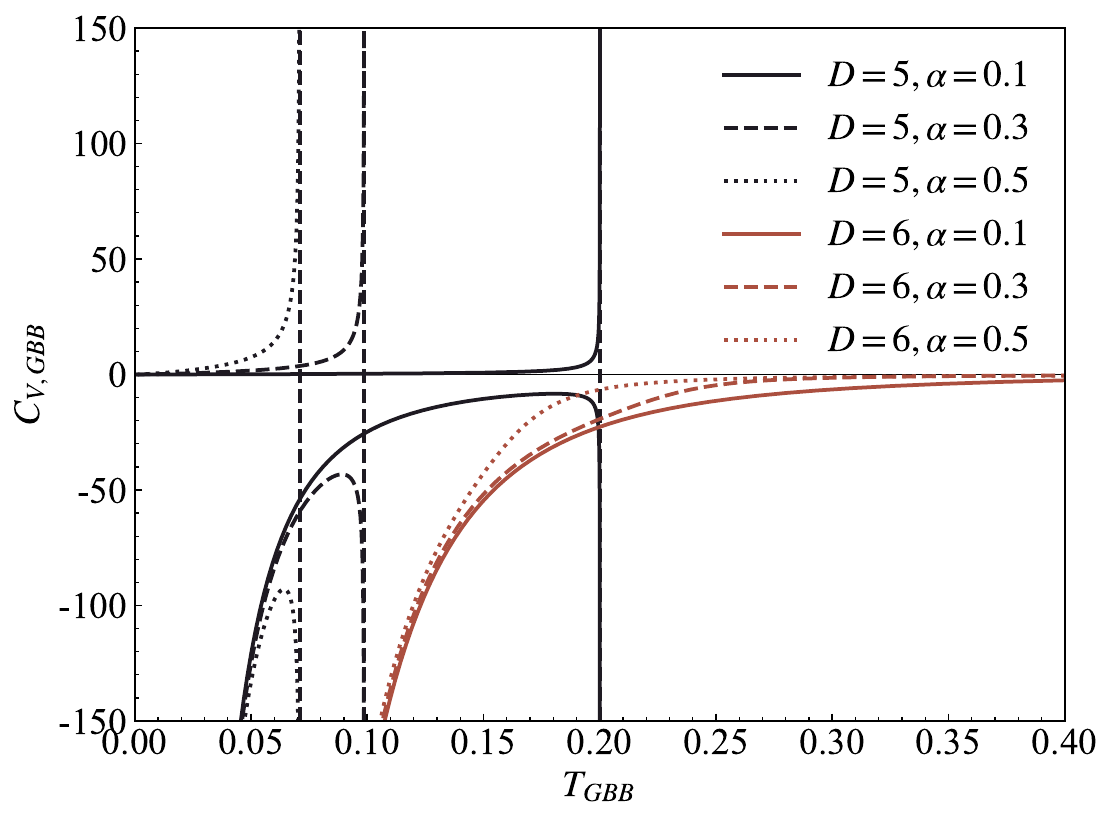}
\caption{The heat capacity $C_{V,GBB}$ versus Hawking temperature for fixed $\Delta=0.5$. Increasing the Gauss-Bonnet coupling $\alpha$ shifts the critical point (divergence) to lower temperatures.}
\label{fig:CV_fix_delta}
\end{figure}

Since the corrected Hawking temperature $T_{GBB}$ remains positive as per Eq.\eqref{temperature_GBB}, the sign of the heat capacity $C_{V,GBB}$ is determined entirely by the slope of the entropy-temperature relation. This behavior is illustrated in Fig.\ref{fig:CV_fix_alpha} and Fig.\ref{fig:CV_fix_delta}.

For the five-dimensional cases (black curves), the heat capacity exhibits a discontinuity at the critical temperature, corresponding to the cusp observed in the entropy plots. As shown in the figures, the heat capacity is positive for the small black hole branch (lower entropy branch), implying that these smaller black holes are thermodynamically stable. In contrast, for the relatively large black holes, the slope of the entropy curve becomes negative, resulting in a negative heat capacity. This instability drives the larger black holes to evaporate and eventually disappear. Effectively, the critical temperature acts as a boundary: below this temperature, a stable small black hole remnant can exist.

It is interesting to observe how the parameters affect this stability region. In Fig.\ref{fig:CV_fix_alpha}, we see that increasing the Barrow parameter $\Delta$ shifts the divergence point to higher temperatures, thereby expanding the domain where stable small black holes can exist. Conversely, Fig.\ref{fig:CV_fix_delta} shows that a stronger Gauss-Bonnet coupling $\alpha$ suppresses the critical temperature, shrinking the stability region.

For higher dimensions ($D=6$, red curves), the story is quite different. As indicated in the figures, and similarly for $D=7$, see Fig.\ref{fig:CV_D7_combined} in Appendix \ref{App.B}, the slope of the Barrow entropy curve remains negative regardless of the strength of the fractal or Gauss-Bonnet corrections. Consequently, the heat capacity $C_{V,GBB}$ is strictly negative throughout the entire parameter space. This confirms that six- or seven-dimensional Gauss-Bonnet black holes are intrinsically unstable and will inevitably exhaust their energy through radiation.

%%%%%%%%%%%%%%%%%%%%%%%%%%%%%%%%%%%%%%%%%%%%%%%%%%%%%%%%%%%%%%%%
%%%%%%%%%%%%%%%%%%%%%%%%%%%%%%%%%%%%%%%%%%%%%%%%%%%%%%%%%%%%%%%%
%%%%%%%%%%%%%%%%%%%%%%%%%%%%%%%%%%%%%%%%%%%%%%%%%%%%%%%%%%%%%%%%

\section{Conclusion and Discussion}

In this work, we carried out a thorough thermodynamic study of $D$-dimensional Gauss-Bonnet black holes modified by Barrow entropy. Specifically, we explored the interplay between the Barrow fractal parameter $\Delta$ and the Gauss-Bonnet coupling $\alpha$. Numerically, the results show a clear dependence on dimensionality. For $D=5$, the stabilizing effect of the Barrow entropy correction shifts the phase transition critical point to higher temperatures. In contrast, for $D\ge6$, the system remains intrinsically unstable with negative heat capacity regardless of the parameter strength.

These findings offer significant insights into quantum gravity phenomenology, particularly regarding the geometric nature of black hole entropy. Unlike standard logarithmic corrections that arise from thermal fluctuations \cite{mahapatra2018logarithmic} or quantum scattering amplitudes in Einstein-scalar-Gauss-Bonnet models \cite{latosh2024black}, the Barrow entropy encodes the non-perturbative fractal structure of the spacetime foam \cite{abreu2025surface, atazadeh2021maximum}. This geometric origin ties deeply into the unified laws of emergence in Lovelock gravity \cite{hassan2023unified, kunstatter2012geometrodynamics}, which fundamentally distinguishes it from information-theoretic generalizations like Tsallis or Rényi entropies \cite{jawad2025geometrothermodynamics, gashti2025non, afshar2025topological, iqbal2018thermodynamics}.

The difference between geometric and statistical corrections leads to diverse thermodynamic outcomes. While Rényi statistics can induce phase transitions in AdS \cite{samart2023ads} or even asymptotically flat spacetimes \cite{bhattacharjee2025thermodynamics}, and affect the topological classes of black holes \cite{qolibikloo2019more, pastras2014charged}, our numerical results in Figs.\ref{fig:ST_fix_alpha} and \ref{fig:ST_fix_delta} indicate that the geometric Barrow parameter $\Delta$ acts more locally. It successfully stabilizes $D=5$ solutions, as evidenced by the characteristic cusp and positive slope branch in the $S-T$ diagrams shown by the black curves in Fig.\ref{fig:ST_fix_alpha}, similar to effects seen in non-extensive thermal stability studies \cite{mushtaq2025non}. However, this geometric deformation is not enough to tame the intrinsic instability of $D \ge 6$ flat Gauss-Bonnet black holes, which is clear from the monotonic behavior of the red curves in Fig. \ref{fig:ST_fix_alpha}. This aligns with the broader context of dynamical instabilities in 4D Einstein-Gauss-Bonnet gravity \cite{konoplya2020stability} and gravitational radiation evolution \cite{babar2023evolution}, suggesting that high-dimensional stability might require additional mechanisms such as nonlinear electrodynamics \cite{kruglov2021einstein}.

Finally, the stabilizing effect of $\Delta$ in five dimensions, as illustrated in Figs.\ref{fig:T_fix_alpha} and \ref{fig:T_fix_delta}, echoes findings from GUP models where it was observed that GUP corrections in 5D-EGB gravity also lead to stable black hole remnants \cite{pradhan2026thermodynamics}, a result supported by quantum-corrected gas dynamics \cite{li2023effects}. This convergence implies a potentially universal feature, suggesting that whether through spacetime foam or minimal length uncertainty, quantum corrections consistently oppose classical thermal instability.

%%%%%%%%%%%%%%%%%%%%%%%%%%%%%%%%%%%%%%%%%%%%%%%%%%%%%%%%%%%%%%%%
%%%%%%%%%%%%%%%%%%%%%%%%%%%%%%%%%%%%%%%%%%%%%%%%%%%%%%%%%%%%%%%%
%%%%%%%%%%%%%%%%%%%%%%%%%%%%%%%%%%%%%%%%%%%%%%%%%%%%%%%%%%%%%%%%

\appendix 
\setcounter{equation}{0} 
\renewcommand{\theequation}{\thesection.\arabic{equation}}
\section{Explicit Form of Heat Capacity} 
\label{App.A}
Substituting the corrected temperature \eqref{temperature_GBB} and Barrow entropy \eqref{entropy_GBB_2} into Eq.\eqref{capacity}, we derive the explicit expression for the heat capacity of Gauss-Bonnet black holes with Barrow entropy corrections as follows:
\begin{align}
\label{capacity_GBB}
C_{V,GBB}&=-V_{D-2}(D-2)(D+\Delta-2)\left[\alpha(D-5)+(D-3)r_+^2\right]\left(2\alpha r_+^{\frac{D-4}{D-2}\Delta}+r_+^{\Delta+2}\right)\notag\\
&\quad\times\left(2\alpha r_+^{\frac{D-4}{D-2}\Delta+D-2}+r_+^{D+\Delta }\right)\Big\{8\alpha^2(D-5)[(D-4)\Delta+D-2]r_+^{\frac{D-4}{D-2}\Delta+2}\notag\\
&\qquad+8\alpha(D-3)[D(\Delta -1)-4\Delta+2]r_+^{\frac{D-4}{D-2}\Delta+4}\notag\\
&\qquad+4\alpha(D-5)(D-2)(\Delta+3)r_+^{\Delta+4}\notag\\
&\qquad+4(D-3)(D-2)(\Delta +1) r_+^{\Delta +6}
\Big\}^{-1}.
\end{align}

\section{Thermodynamic Behaviors in \texorpdfstring{$D=7$}{D=7}}
\label{App.B}
\setcounter{figure}{0}
\renewcommand{\thefigure}{B\arabic{figure}}

In this appendix, we present the thermodynamic quantities for the seven-dimensional Gauss-Bonnet black holes ($D=7$). As discussed in the main text, the behaviors are qualitatively similar to the $D=6$ case, exhibiting no phase transitions and monotonic instability.

\begin{figure}[tp]
\centering
\begin{minipage}[b]{0.48\textwidth}
\centering
\includegraphics[width=\textwidth]{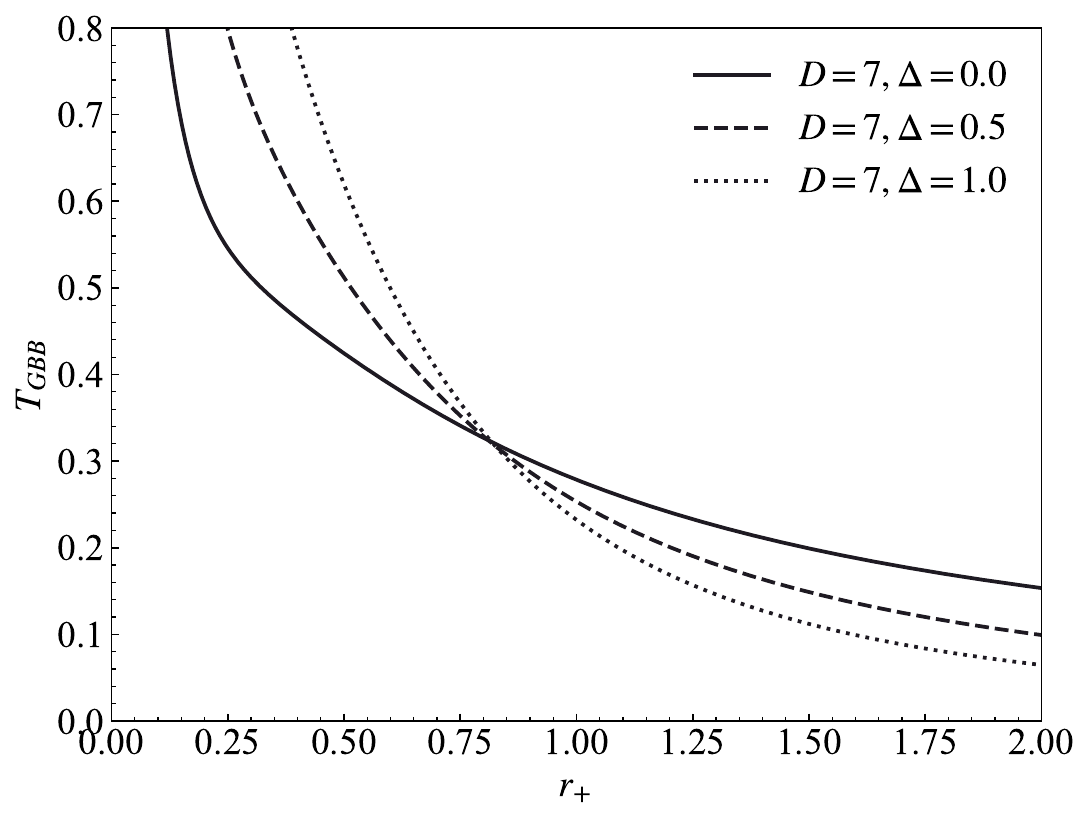}
\caption*{(a) Fixed $\alpha=0.1$}
\end{minipage}
\hfill
\begin{minipage}[b]{0.48\textwidth}
\centering
\includegraphics[width=\textwidth]{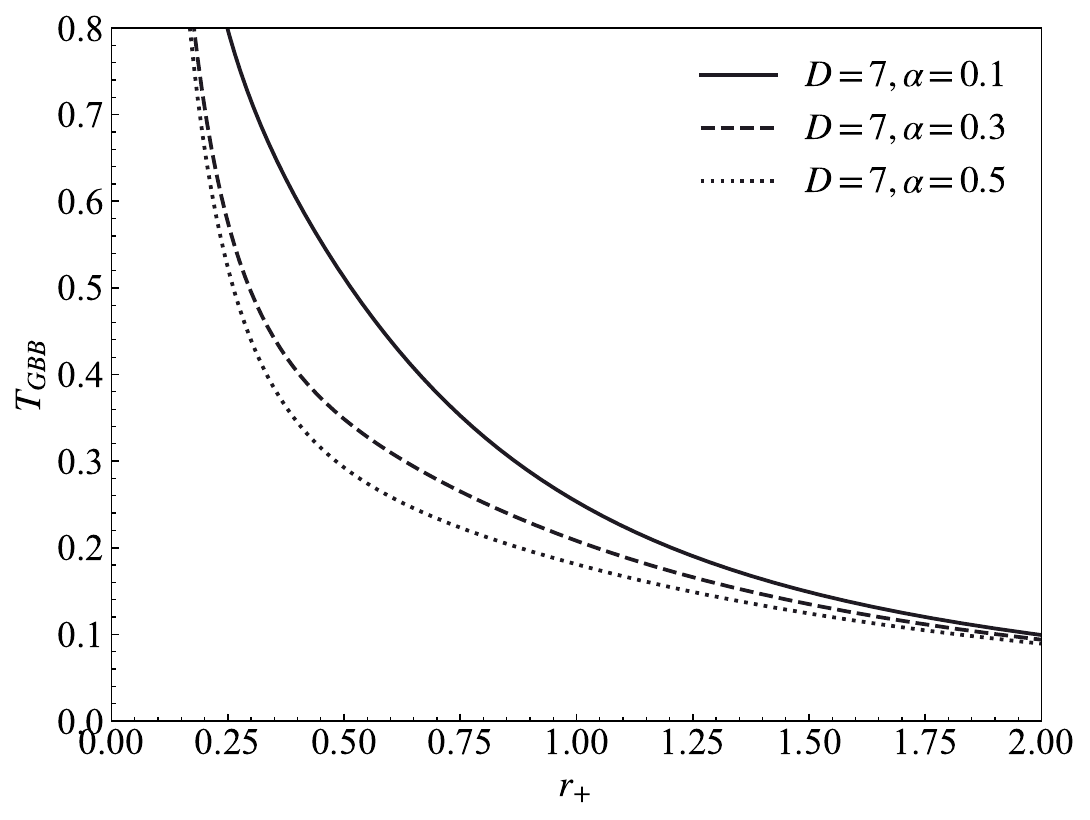}
\caption*{(b) Fixed $\Delta=0.5$}
\end{minipage}
\caption{The Hawking temperature $T_{GBB}$ as a function of horizon radius $r_+$ for $D=7$.}
\label{fig:Tr_D7_combined}
\end{figure}

\begin{figure}[tp]
\centering
\begin{minipage}[b]{0.48\textwidth}
\centering
\includegraphics[width=\textwidth]{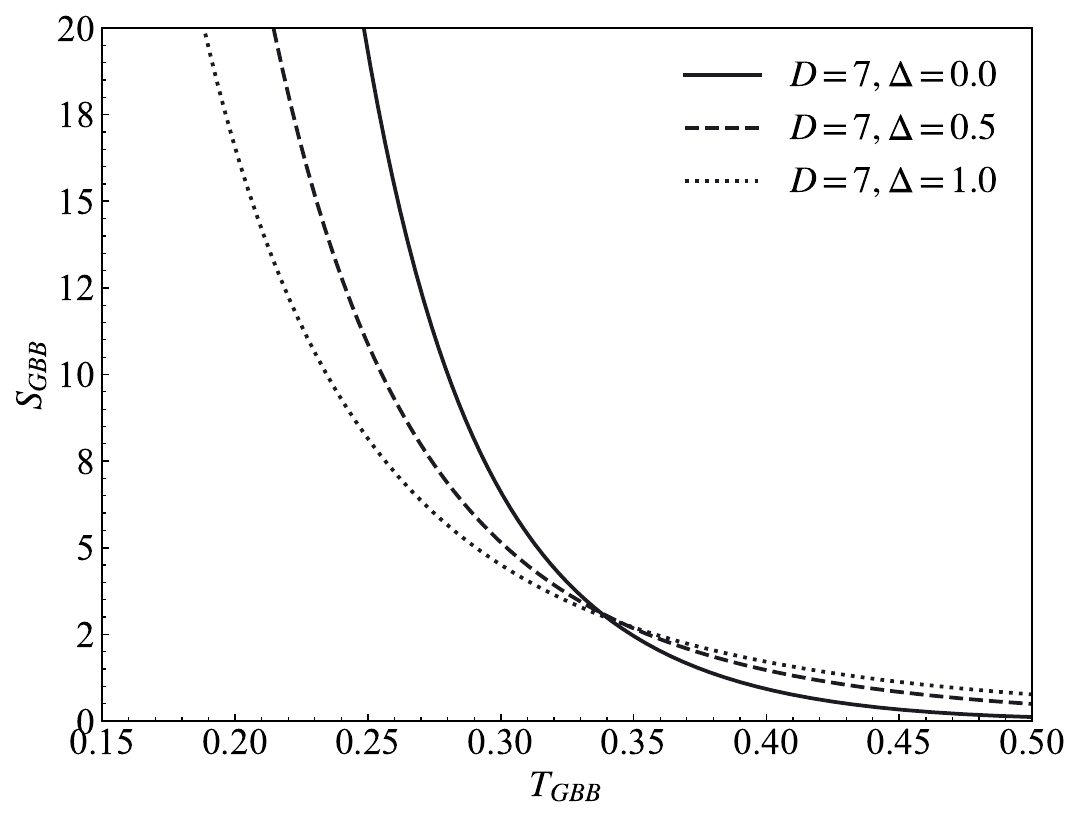}
\caption*{(a) Fixed $\alpha=0.1$}
\end{minipage}
\hfill
\begin{minipage}[b]{0.48\textwidth}
\centering
\includegraphics[width=\textwidth]{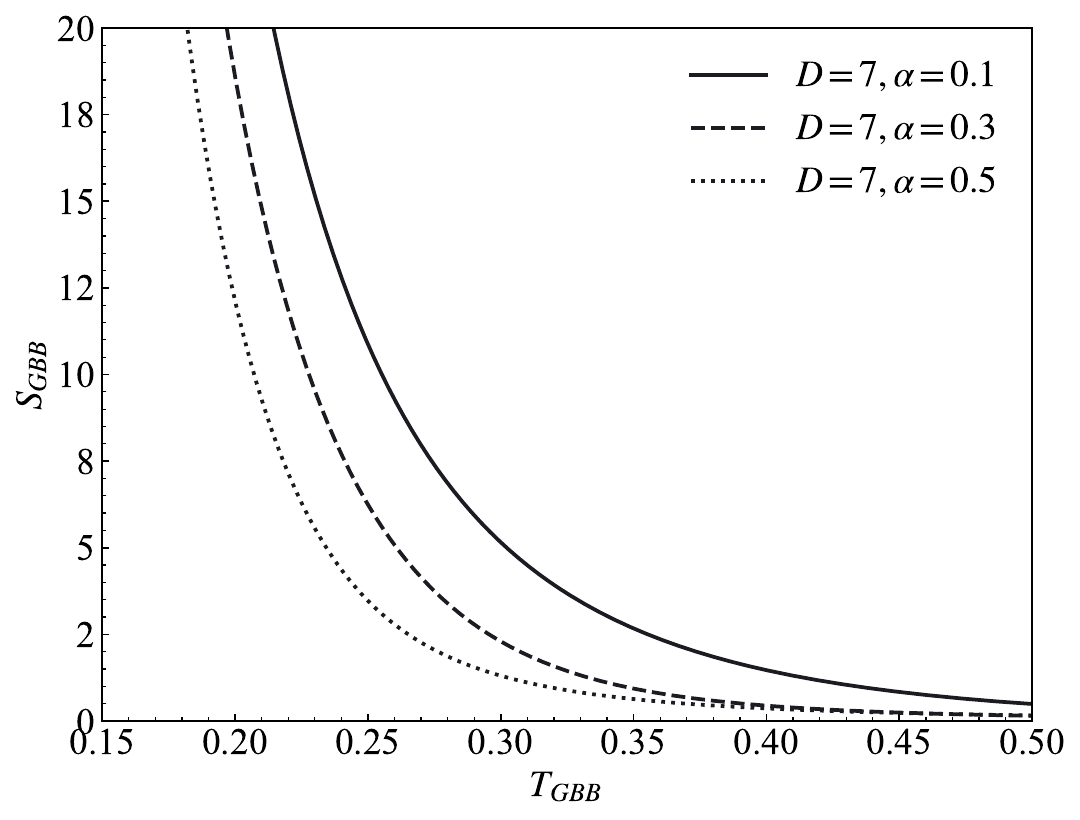}
\caption*{(b) Fixed $\Delta=0.5$}
\end{minipage}
\caption{The Entropy-Temperature ($S_{GBB}-T_{GBB}$) relations for $D=7$.}
\label{fig:ST_D7_combined}
\end{figure}

\begin{figure}[tp]
\centering
\begin{minipage}[b]{0.48\textwidth}
\centering
\includegraphics[width=\textwidth]{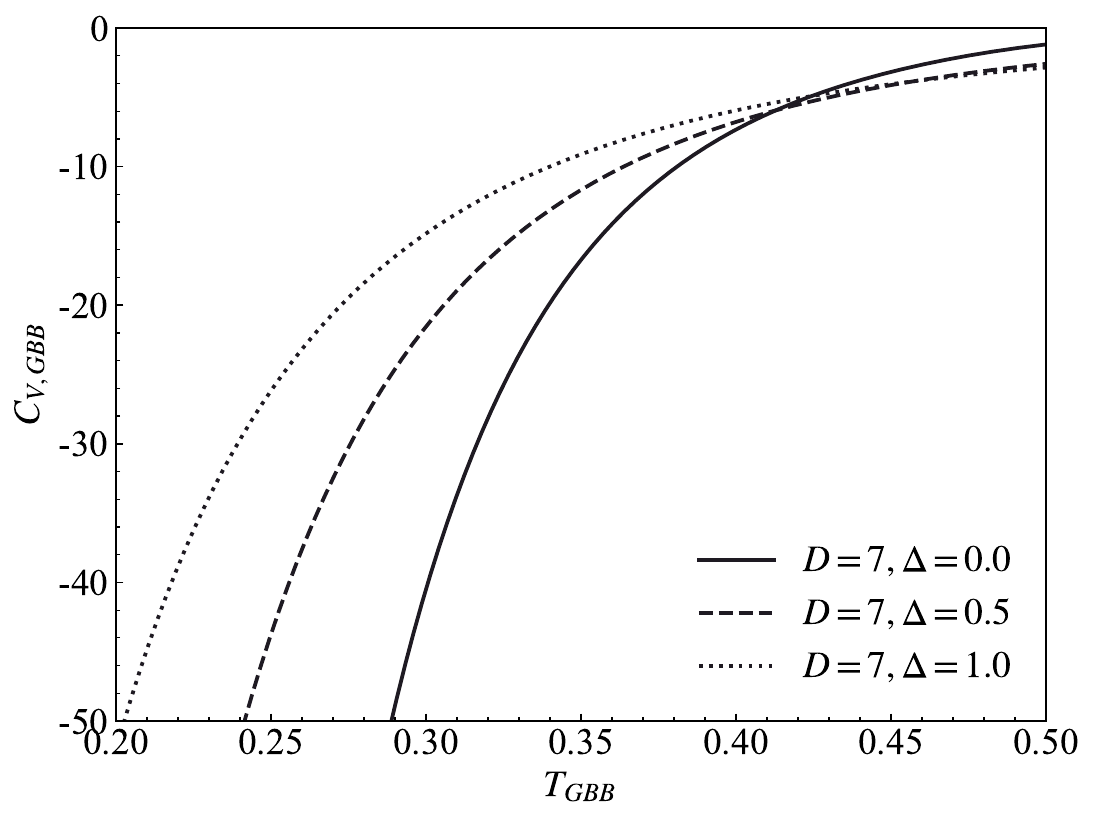}
\caption*{ (a) Fixed $\alpha=0.1$}
\end{minipage}
\hfill
\begin{minipage}[b]{0.48\textwidth}
\centering
\includegraphics[width=\textwidth]{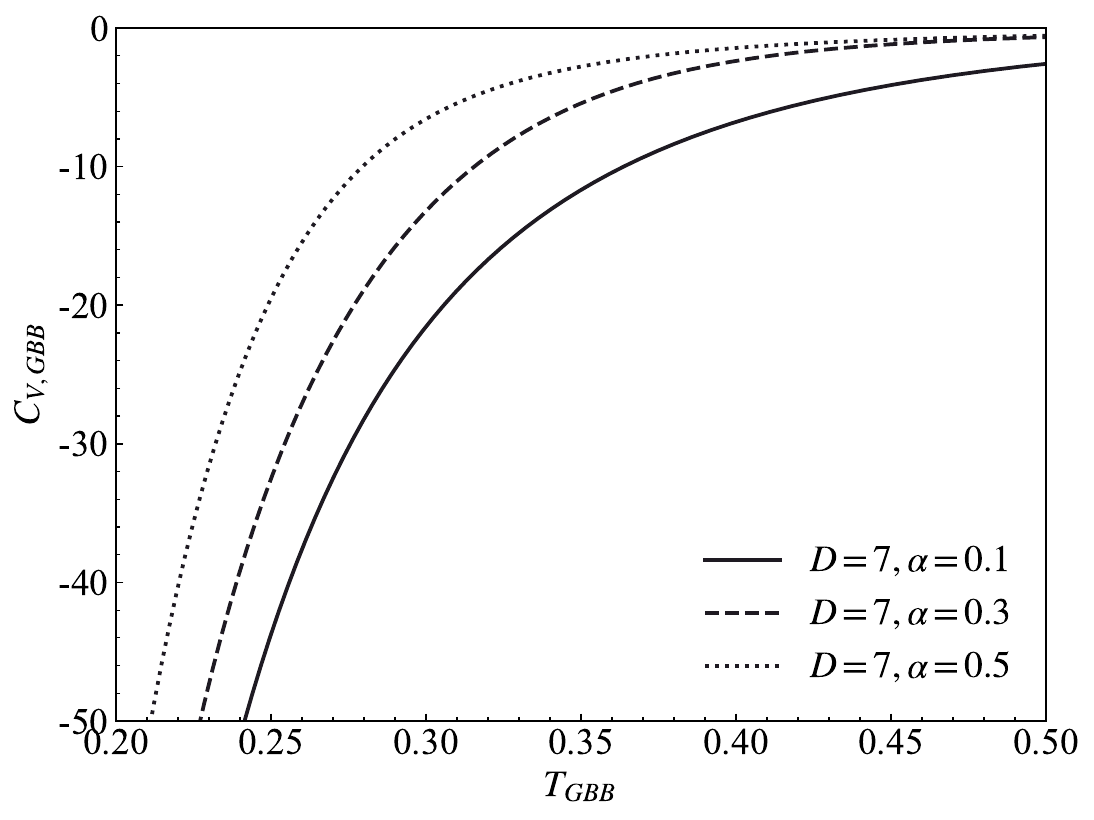}
\caption*{ (b) Fixed $\Delta=0.5$}
\end{minipage}
\caption{The heat capacity $C_{V,GBB}$ for $D=7$ black holes.}
\label{fig:CV_D7_combined}
\end{figure}

\vspace{1cm}
\noindent \textbf{Acknowledge}

This work is partly supported by the Shanghai Key Laboratory of
Astrophysics 18DZ2271600.

\bibliographystyle{unsrt}
\bibliography{main}

\end{document}